\documentclass{article}

 \PassOptionsToPackage{numbers, compress}{natbib}

\usepackage[utf8]{inputenc} 
\usepackage[T1]{fontenc}    
\usepackage{hyperref}       
\usepackage{xurl}
\usepackage{booktabs}       
\usepackage{amsfonts}       
\usepackage{nicefrac}       
\usepackage{microtype}      
\usepackage{graphicx}
\usepackage{placeins}
\usepackage{chngcntr}
\usepackage{placeins}
\usepackage{tabto}
\usepackage{natbib}
\usepackage{breakcites}
\usepackage[a4paper, total={6in, 8in}]{geometry}
\usepackage{wrapfig}
\usepackage{multicol}
\usepackage{listings}
\usepackage{float}
\usepackage{multirow}
\usepackage{amsmath}

\usepackage{xcolor}
\usepackage[inline,shortlabels]{enumitem}
\usepackage{capt-of}

\usepackage{rotating}
\date{}
\title{Ethical Considerations and Statistical Analysis of Industry Involvement in Machine Learning Research}

\author{%
 	Thilo Hagendorff\footnote{Alphabetical ordered. Authors contributed equally.}\\
 University of Tübingen \\
Cluster of Excellence Machine Learning\\
Tübingen, Germany \\
	\texttt{thilo.hagendorff@uni-tuebingen.de} \\
 \\ 
  	Kristof Meding\(^*\)\\
 University of Tübingen \\
Neural Information Processing Group \\
	Tübingen, Germany \\
	\texttt{kristof.meding@uni-tuebingen.de} \\
}

\begin{document}

\maketitle
\begin{abstract}
Industry involvement in the machine learning (ML) community seems to be increasing. However, the quantitative scale and ethical implications of this influence are rather unknown. For this purpose, we have not only carried out an informed ethical analysis of the field, but have inspected all papers of the main ML conferences NeurIPS, CVPR, and ICML of the last 5 years---almost 11,000 papers in total. Our statistical approach focuses on conflicts of interest, innovation and gender equality. We have obtained four main findings: (1) Academic-corporate collaborations are growing in numbers. At the same time, we found that conflicts of interest are rarely disclosed. (2) Industry publishes papers about trending ML topics on average two years earlier than academia does. (3) Industry papers are not lagging behind academic papers in regard to social impact considerations. (4) Finally, we demonstrate that industrial papers fall short of their academic counterparts with respect to the ratio of gender diversity. We believe that this work is a starting point for an informed debate within and outside of the ML community.
\end{abstract}

\section{Introduction}
The number of papers submitted and accepted at the major machine learning (ML) conferences is growing rapidly. Besides submissions from academia, big tech companies like Amazon, Apple, Google, and Microsoft submit a large number of papers. But the influence of these companies on science is unclear. Do they drive innovation? What are potential upsides and downsides of industry involvement in ML research? What are the possible ramifications of conflicts of interest? In order to investigate these topics, namely the industry involvement in ML research and its associated ramifications that range from questions about conflicts of interests, to scientific progress, research agendas, and gender balance. We have conducted a statistical data analysis of the field. Our analysis serves to answer four overarching research questions. First of all, we will develop a quantitative analysis of the proportion of industry, academic, and academic-corporate collaboration papers within the three major ML conferences, namely the Conference and Workshop on Neural Information Processing Systems (NeurIPS), International Conference on Machine Learning (ICML), and Conference on Computer Vision and Pattern Recognition (CVPR). Secondly, we aim to find out whether conflicts of interest are disclosed in those cases in which they are pertinent. Answering these questions will be of importance in order to assess potential changes in conference policies on transparency statements and to inform discourses on "AI governance" \cite{Daly.2019}. Thirdly, we are interested in the role industry papers play with regard to scientific progress and ethical concerns, as well as whether they are, in this respect, any different to academic research. And finally, we investigate gender balances, particularly with regard to proportions of women working on industry papers. We also discuss our findings in light of recent ethical research and implications for the ML community \cite{Mittelstadt.2019}. In the following paragraphs, we give a theoretical introduction into the mentioned issues and discuss their ethical implications.

\subsection{The ethics of industry funding and conflicts of interest}
A concern connected to industry funding is that research agendas are skewed. More applied topics and short-term benefits are favored over basic science and its potential long-term outcomes. This causes research strands to strongly orientate towards corporate interests \cite{Washburn.2008}, or, more severely, to the plain distortion or suppression of certain research results in order to produce favorable outcomes for the respective sponsor. This is called “industry bias” \cite{Lundh.2017, Probst.2016, Krimsky.2013}. This bias can occur due to payments for services, the commodification of intellectual property rights, research funding, job offers, startups or companies owned by scientists, consultation opportunities, and the like. Especially criticised is the fact that machine learning conferences are hardy ever free of industry sponsors \cite{Abdalla.2020}. These sponsors can then control the conference's agendas and the questions being asked there.
In order to delve further into the subject, we will examine conflicts of interest which are a common side effect of industry involvement in academic research in general \cite{Etzkowitz.2000, DEste.2007, Boardman.2009, Bruneel.2010}. A substantial amount of literature is dedicated to reflecting on conflicts of interest that can occur in clinical practice, education, or research \cite{Rodwin.1993, Fickweiler.2017, Thompson.1993}. As a consequence of conflicts of interest in research, medical journals require researchers to name funding sources. The public disclosure of funding sources, affiliations, memberships etc. are supposed to inform those who receive scientific information or advice in order to fill information gaps. This allows them to assess the information or advice to its full extent.
But what exactly are conflicts of interest? While it is hard to find a universal definition, a common denominator is that conflicts of interest arise when personal interests interfere with requirements of institutional roles or professional responsibilities \cite{Komesaroff.2019}. Here, interests can be seen as goals that are aligned with certain financial or non-financial values that have a particular, maybe detrimental effect on decision-making. Coexistence of conflicting interests results in an incompatibility of two or more lines of actions. In modern research settings, dynamic and complex constellations of conflicting interests frequently occur \cite{Komesaroff.2019}. For instance, conflicts of interest do not only pose a problem in cases where researchers intentionally follow particular interests that undermine others. Many effects arising from conflicts of interest take effect on subconscious levels \cite{Cain.2008, Moore.2004b, Dana.2003}, where actions are rationalized by post hoc explanations \cite{Haidt.2001}. Many studies, especially in the field of medical research, show that even when physicians report that they are not biased by financial incentives, they actually are \cite{Orlowski.1992, Avorn.1982}. This means that despite researchers' belief in their own integrity and the idea that financial opportunities, honorariums, grants, awards, or gifts have no influence in their line of action, opinion, or advice, the influence is, in fact, measurable.
Psychological research has shown that individuals often succumb to various biases that steer their behavior \cite{Chavalarias.2010,Ioannidis.2005,Kahneman.2012b,Tversky.1974}. These are so-called “self-serving biases”, meaning that fairness criteria, assumptions about the susceptibility towards conflicts of interest, or other ways of evaluating issues are skewed towards one’s own favor \cite{McKinney.1990}. One famous self-serving bias is exemplified by the fact that physicians assume that small gifts do not significantly influence their behavior, while actually, the opposite is true \cite{Brennan2006}. Even small favors elicit the reciprocity principle, meaning that there is a clear influence or bias on an individual’s behavior. These biases are not necessarily associated with lacking moral integrity or even corruptibility. On the contrary, they can be assigned to an “ecological rationality”, meaning that an individual’s behavior is adapted to environmental structures and certain cognitive strategies \cite{Arkes.2016, Gigerenzer.2001}. Nevertheless, conflicts of interest can have or actually do have dysfunctional effects on the scientific process. Hence, the scientific community does well in finding a way to deal with them properly. This is mostly done by obliging researchers to disclose conflicts of interest. While this is an accepted method in many scientific fields, it can actually have negative effects. These so-called ``perverse effects'' are described by Cain et al. \cite{Cain.2005, Crawford.1982}. On the one hand, Cain and colleagues demonstrate that disclosing conflicts of interest does not lead people to relativize claims by biased experts sufficiently since disclosure can in some cases increase rather than decrease trust. On the other hand, and more importantly, experts who reveal conflicting interests may thus feel free to exaggerate their advice and claims since they have lowered their guilty conscience about spreading misleading or biased information. While transparency statements have side effects, they should certainly not be omitted entirely.

\subsection{Drivers of scientific progress and innovation}
Despite the manifold pitfalls that are caused by the intermingling of academia and industry, studies show that particularly corporate-sponsored research can be very valuable for science itself as well as for society as a whole \cite{Wright.2014}. Hence, one has to discuss another concomitant of industry involvement in research, namely industry's potential innovative strength. Industry involvement in the sciences can not only provide more jobs, lead to tangible applications of scientific insights, provide life-enhancing products, increase a society's wealth, but also lead to much-cited papers, and spur innovations. Researchers \cite{Wright.2014} have shown that corporate-sponsored inventions resulted in licenses and patents more frequently than federally sponsored ones---although this alone does not mean that industry is more innovative per se. Current research also shows that machine learning research in the private sector tends to be less diverse topic-wise than research in academia \cite{Klinger.2020}. Furthermore, corporations are often seeking university partners in order to widen their portfolio of products, business models, and profit opportunities. This can nudge academic partners to act progressively, towards novel, unprecedented experiments, research ideas, and speculative approaches \cite{Evans.2010}. On detours, industry funds lead to scientific progress and innovation. Research on innovation processes has shown that organizations are typically not innovating internally, but in networks, in social relationships between members of different organizations, in technology transfer offices, science parks, and many other university-industry collaborations \cite{Perkmann.2007}. These collaborations can emerge via research papers, conferences, meetings, informal information exchange, consulting, contract research, hired graduates, a joint work on patents or licences, etc., and play a vital role in driving innovation processes \cite{Cohen.2002b}. All in all, scientists' sensitivity towards opportunities of industry funding may in fact cause ``deformed'' research agenda settings. This does not necessarily mean, though, that innovations, scientific progress, and its positive effects for society are diminished. With our data analysis we aspire to find out how this constellation is reflected in the field of ML research.
Academic engagement, i.e. the involvement of researchers in university-industry knowledge transfer processes of all kinds, is a common by-product of academic success. Scientists who are well established, more senior, have more social capital, more publications, and more government grants, are at the same time more likely to have industrial collaborators \cite{Perkmann.2013}. This is due to the “Matthew effect”, meaning that researchers who are already successful in their field of research are more likely to reinforce this success with industry engagements whose returns continuously lead to more academic success. Researchers involved in commercialization activities publish more papers in comparison to their non-patenting colleagues \cite{Fabrizio.2008,Breschi.2007}. Scientific success in ML research seems to go hand in hand with industry collaborations. However, industry-driven research or research that is intended to be commodified is, in most cases, more secretive and less accessible for the public.
Taking all these considerations into account, a further objective of our data analysis is to scrutinize the innovative strength of industry research. For that purpose, we will not only conduct a citation analysis, but also look at concrete machine learning methods and measure whether industry or academic papers pushed these methods before they became a commonly used standard tool for machine learning practitioners. Lastly, we look at social impact awareness in industry research and also compare it to academic research.

\subsection{Statistics on gender imbalances}
The final issue we are going to investigate is that of gender aspects and their entanglement with industry research. Noticeably, male academics are significantly more likely to have industry partners than female scientists \cite{Perkmann.2013}. This finding corresponds to the fact that ML research has a diversity imbalance, indicating that male researchers strikingly outnumber females. Statistics show that only a small share of authors at major conferences are women. The same holds true for the proportion of ML professors, the affliction of tech companies with heavy gender imbalances, women's tendency to leave the technology sector, as well as the fact that they are payed less than men \cite{MyersWest.2019, Simonite.2018}. Further diversity dimensions like ethnicity, intersexuality, and many other minorities or marginalized groups are often not statistically documented. Tech companies have even thwarted access to diversity figures to avoid public attention on the under-representation of women and minorities \cite{Pepitone.2013}. All in all, the ``gender problem'' of the ML sector does not only manifest itself on the level of lacking workforce diversity, but in the functionality of software applications, too \cite{Leavy.2018}. Despite these rather general observations and statistics, we want to find out whether gender imbalances have a particularly pronounced manifestation in the context of industry ML research. Inspired by previous research on gender imbalances \cite{Andersen.2019}, we scrutinise the ratio of female (last) authors in academia and industry papers. This is of importance to prove or disprove common intuitions about the disadvantage against women, which is allegedly or actually stronger in companies compared to university contexts.
\vspace{-0.4cm}
\section{Methods}
\subsection{Analysing 10,772 ML papers}
At this point, we will briefly describe the methods we have used to conduct our statistical analysis. More detailed information about the process can be found in the supplementary material. All in all, our analysis focuses on three major ML conferences: ICML, CVPR, and NeurIPS. We downloaded all articles available spanning the years 2015 to 2019 from the respective conference proceedings. Altogether, the data set contains 10,807 papers. The papers were downloaded using the python-tool Beautiful Soup (v. 4.8.2). We extracted the text with pdftotext (v. 0.62.0) and analyzed the text with a self-written script. Using this method, we were able to search 10,772 papers (99.7\%). Some of the papers are, for example, not searchable because their text is embedded as an image. We are not only analysing the papers themselves, we are also interested in the metadata, namely affiliations and authors. For the analysis of the affiliations, we extracted them from the texts and categorised them into academic and industry affiliations. We define a paper as academic if it contains one of the following terms (see supplementary material \ref{sec::affTerms} for more information on why we use these terms only):\\
\\
California Institute of Technology
/ Ecole
/ EPFL
/ ETH Z\"urich
/ INRIA
/ Kaist
/ Massachusetts Institute of Technology
/ MILA
/ MIT
/ ParisTech
/ Planck
/ RIKEN
/ TU Darmstadt
/ Université
/ Universiteit
/ University. \\
\\
For the definition of a paper as industry we use the following terms: \\
\\
Adobe
/ AITRICS
/ Alibaba
/ Amazon
/ Ant Financial
/ Apple
/ Bell Labs
/ Bosch
/ Criteo
/\linebreak Data61
/ DeepMind
/ Expedia
/ Facebook
/ Google
/ Huawei
/ IBM
/ Intel
/ Kwai
/ Microsoft
/ NEC
/ Netflix
/ NTT
/ Nvidia.
/ Petuum
/ Qualcomm
/ Salesforce
/ Siemens
/ Tencent
/ Toyota
/ Uber
/ Vector Institute
/ Xerox
/ Yahoo
/ Yandex.\\
\\
Unless otherwise stated, we define a paper as academic if it does not contain an industry term in the affiliation section and a paper belonging to industry if it does not contain an academic term. A paper is defined as mixed if contains an academic and an industry affiliation. In total, 90.2\% of all papers contain at least one of the terms from academia or industry listed above. These numbers are entirely dependent on the fact that the authors actually declare all their affiliations in the paper. Furthermore, we extracted the author's names and the titles of the papers directly from the websites, not from pdf documents. For this purpose, we once again used Beautiful Soap. We extracted 41,939 authors. However, many authors have multiple accepted papers, and thus, the number of authors is reduced to 18,060 unique authors. All information in text and graphics about the number of authors refers to these unique authors. The genders of the names were determined using the name-to-gender service GenderAPI. GenderAPI offers the highest accuracy of the name-to-gender tools \cite{Santamaria2018} and was able to determine the gender of 17,412 authors (96.4\%). Finally, we downloaded the citations received for each individual paper using the Microsoft Academic Knowledge API \cite{Sinha2015} (date citations received: 03.29.2020). This was successful for 10,616 papers (98.2\%). 
Our approach has three (possible) limitations. Firstly, our results should be understood as general and robust trends but not as exact numbers, since it is not is possible to extract data from the papers in all cases. A further limitation of our method that is particularly relevant for our analysis of conflicts of interest is that we cannot detect cases where researchers have academic and industry affiliations at the same time but state only one of them in the respective research paper. Moreover, we would like to point out that the data set is smaller for the industry analysis. Small data sets tend to produce extreme results---in both positive and negative directions. Nevertheless, we believe that this is not a problem in our case as our results, as we will see in the following section, are very robust.

\subsection{Error bars and statistical modelling}
Here, we briefly describe the way we calculated error bars and explain our statistical model approach. All details can be found in \ref{sec:ModelDetails}.\\
In most of the analysis we extract ratios which follow a binomial distribution. Following this, we used the Wilson-Score approximation of a binomial distribution for the confidence intervals of individual data points (figure 1b, 2b, 3b, 3c, 4a, and 4b). For the calculation of confidence intervals of median citation counts a 1000-fold bootstrap approach was used. This approach is less influenced by the underlying data distribution compared to a parametric approach. \\
However, these confidence intervals are not suitable for the comparison of different data points in our figures, in this case, academic, industry, and mixed papers. For the comparison of academia and industry as well as trends in time, a (generalized) linear model approach was used \cite{faraway2014linear}. We used the metafor-package (v. 2.4-0) in R for all figures except for figure 3a. The metafor-package allows us to include the uncertainty of data points. In figure 3a we used the negative binomial fitting procedure \textit{glm.nb} from the MASS-package (v. 7.3-51.6), since we are able to obtain an individual citation count to every individual paper.

\section{Results}
\subsection{Subtle conflicts of interest in academia}
\begin{figure}[h!]
	\centering
	\includegraphics[width=1\textwidth]{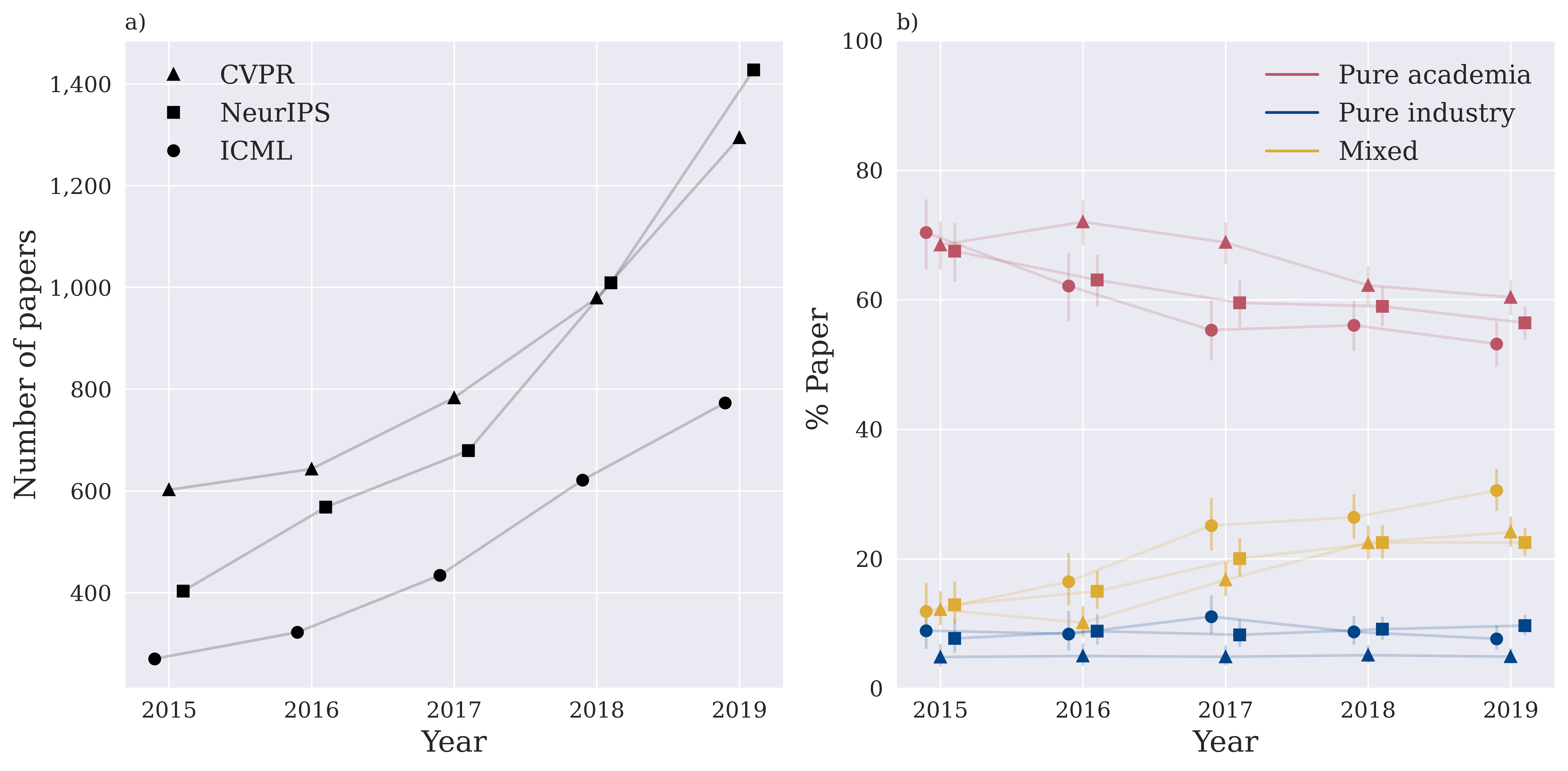}
	\caption{(a) Progression of the number of papers at major ML conferences (a) and (b) institutional affiliations. Please note the numbers in (b) do not add up to 100\% because we were only able to extract this information for 90\% of the papers, see methods and supplementary information. Error bars in this example and all following figures illustrate a 95\% confidence interval.}
	\label{fig:Fig1}
\end{figure}

Figure \ref{fig:Fig1}a plots the number of papers accepted at ICML, NeurIPS, and CVPR between 2015 - 2019. The number of accepted papers is steadily increasing. Figure \ref{fig:Fig1}b shows whether the paper includes authors with affiliations from academia, industry, or both. While the ratio of industry papers is stable, an increasing ratio of papers have affiliations from both academia and industry. We obtain with our linear analysis a slope of -2.9\%/year (95\%-\(CI\): \([-3.7,-2.1]\), \(p < 10^{-5}\)) for academia,  0.1\%/year (95\%-\(CI\): \([-0.3, 0.4]\), \(p = 0.70707\)) for industry papers and 3.6\%/year (95\%-\(CI\): \( [2.7,4.4]\), \(p < 10^{-5}\)) for mixed papers.

\begin{figure}[]
		\centering
		\includegraphics[width=0.7\textwidth]{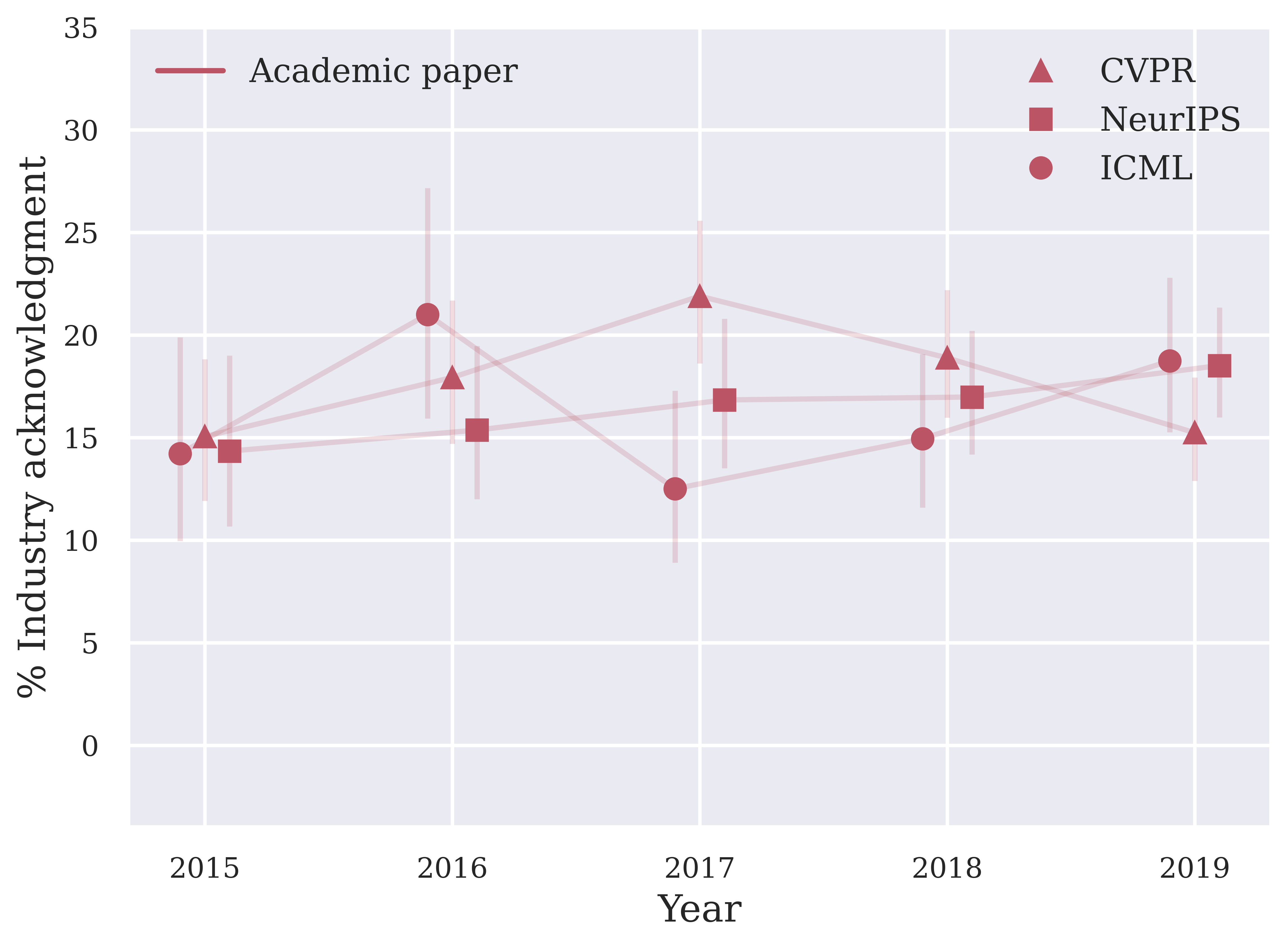}
		\caption{Percentage of paper from academia which contain an industry acknowledgement.}
		\label{fig:Fig2}
	\end{figure}

Furthermore, we extracted the acknowledgements of all papers from academia and searched them for terms of industry affiliations (Google, Facebook etc.). This gives us an insight into whether academic papers acknowledge industry funding, grants etc. In fact, we calculated the conditional probability \(p(\text{industry acknowledgement}|\text{academia})\). 
With recourse to the insights from figure \ref{fig:Fig1}b, there is no doubt that purely academic papers make up the largest part of submissions to all major ML conferences, not industry papers.
\newpage
However, figure \ref{fig:Fig2} shows the ratio of purely academic papers with industry acknowledgements. Roughly 15\% - 20\% of purely academic papers contain an industry acknowledgement. No significant trend in time is found (0.5\%/year, 95\%-\(CI\): \([-0.5,1.5]\),\(p = 0.32382\)). Finally, we also searched for the terms ``conflict of interest''---the plural ``conflicts of interest'', which did not lead to a single finding--- and ``disclosure'' in order to identify whether such influences are named. Only three of more than ten thousand papers contain an explicit conflict of interest statement at all. This inquiry shows that on one hand, conflicts of interest are present in many academic research papers, while on the other hand, those conflicts are not clearly stated. This further indicates that it is sensible for ML conferences to demand researchers to add transparency statements to their submissions.\\
All in all, ties between the two social systems \cite{Luhmann.1995b}---university and industry---do seem to become tighter. Academic settings are increasingly mitigating towards corporate tech environments. Moreover, academic papers with no industry affiliations are slightly on the decrease. This urgently calls for an appropriate approach to dealing with conflicts of interests. However, purely academic papers still make up the largest part of the submissions to all major conferences.
	\begin{figure}[!h]
		\centering
		\includegraphics[width=1\textwidth]{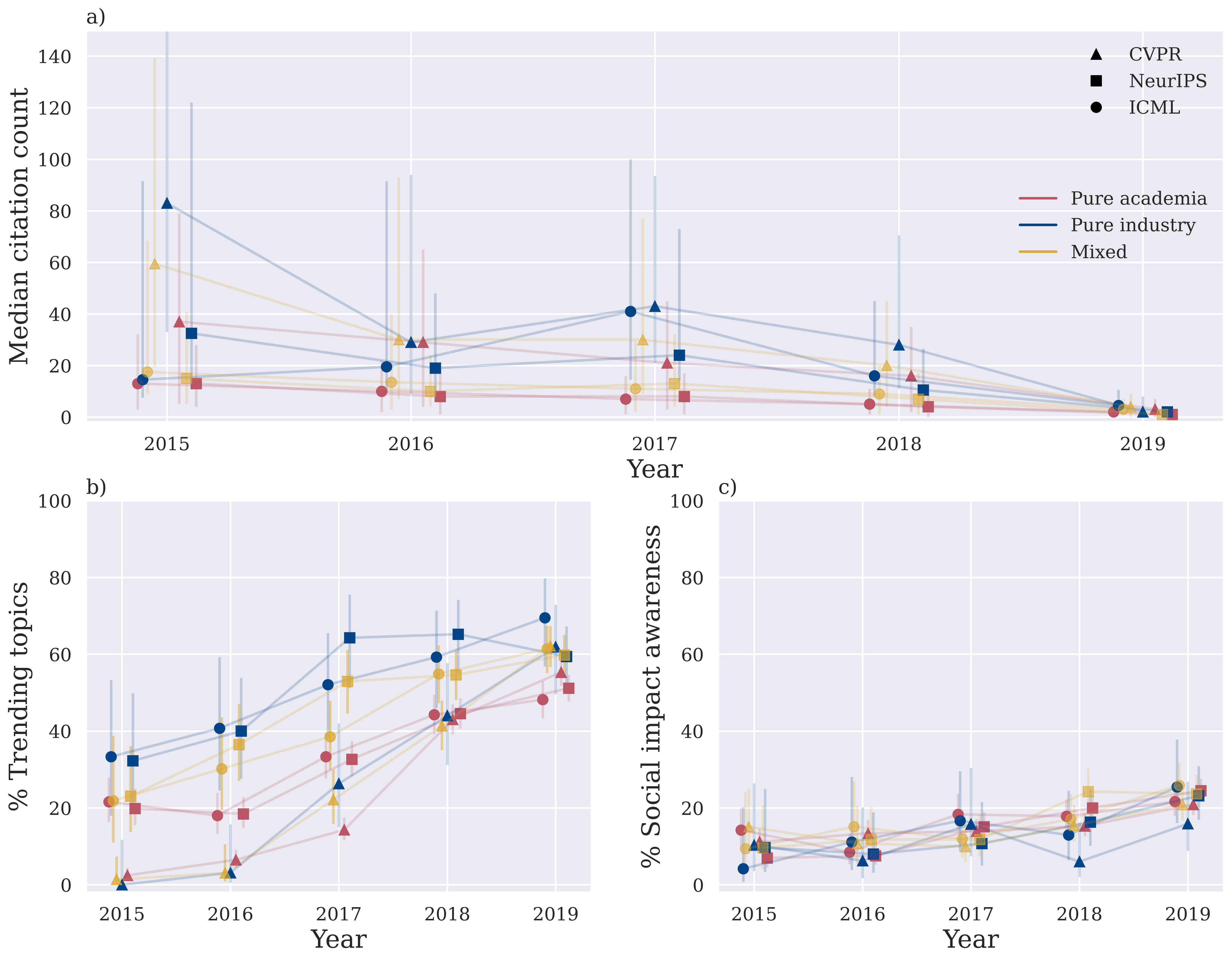}
		\caption{(a) Median number of citations received. (b) Ratio of papers from academia and industry with trending topics: 'adversarial' and 'reinforcement') and (c) papers with social impact terms.}
		\label{fig:Fig3}
	\end{figure}
	
\subsection{Publishing behavior: Industry is setting trends}
Next, we want to find out whether it is academia or industry that propels important parts of the ML field. Thus, we compared industry and academic papers with regard to the average amount of citations they possess. The results are shown in figure \ref{fig:Fig3}a. Our generalized linear model analysis shows that there is a significant difference between academia and industry (estimate = 0.985, 95\%-\(CI\): \([1.1, 0.9]\), \(p < 2\cdot 10^{-16}\)). Due to the negative binomial model, the estimate coefficient has no direct interpretation. \\	While citation analyses are not particularly credible for papers that were published quite recently, since citations are slow to accumulate, citation analyses gain significance over time. Thus, our analysis clearly shows that industry papers from 2015 were cited far more frequently than academic papers. This trend prevails throughout the following years, albeit on a smaller scale.\\
In order to gain further insights into whether it is academia or industry that drives the field of ML, we searched for two terms, the first one being ``adversarial''. This, on the one hand, corresponds to the very popular Generative Adversarial Networks invented 2014 by Goodfellow et al. \cite{Goodfellow2014} and, on the other, to the adversarial attack on neural networks \cite{Szegedy2013}. We also included the term ``reinforcement'' for reinforcement learning. These are topics of increasing interest to the ML field \cite{Biggio2018,Lipton2018}. The results are shown in figure \ref{fig:Fig3}b. The linear model analysis shows a difference of 11.5\% (95\%-\(CI\): \([5.4,17.7]\), \(p < 0.00024\)) between academia and industry. The figure indicates that, with regard to the three mentioned concepts, academia is lagging roughly two years behind industry (ICML and NeurIPS). Similar trends can be found for much more frequently used terms like ``convolution'' and ``deep'' in supplementary figure \ref{fig:FigA1}.\\
In addition, we are interested in whether social aspects are becoming more interesting to the ML community. Thus, we included terms from the social impact category of NeurIPS 2020 (safety, fairness, accountability, transparency, privacy, anonymity, security) and added ``ethic'' as well as ``explainab''. We call these terms social impact terms. Figure \ref{fig:Fig3}c shows the results. Overall, we can see, that the ML community pays more attention to these terms in the course of the past years (3.2\%/year, 95\%-\(CI\): \([2.5,4.0]\), \(p < 10^{-5}\)). But while one may assume that academic papers put a stronger focus on social impact issues in comparison to industry research, this intuition does not hold true. Only a small hardly-significant difference of 3.1\% (95\%-\(CI\): \([0.2,6.0]\), \(p = 0.034\)) between academia and industry is found. The amount of social impact terms is more or less equally shared between academia and industry, showing that ideas of an ethical ``superiority'' of academia do not bear scrutiny.\\
Especially with regard to the results from figure \ref{fig:Fig3}a, we can show that industry papers have higher citation rates compared to academic papers in every year, giving evidence for the high scientific relevance of industry papers. There is no question that industry papers receive greater attention from the scientific community than academic papers. A confound of this analysis is that one may assume that academic researchers, who are strikingly successful, are likely to be hired by ML companies, which then cause industry papers to have more citations on average than academic papers. Thus, it is difficult to state whether industry research has more scientific impact because of the industry context itself or because of companies' strategic hiring policies and the corresponding migration of successful university researchers to companies.\\

\subsection{Gender equality: Industry is behind academia}

\begin{figure}[h!]
		\centering
		\includegraphics[width=0.9\textwidth]{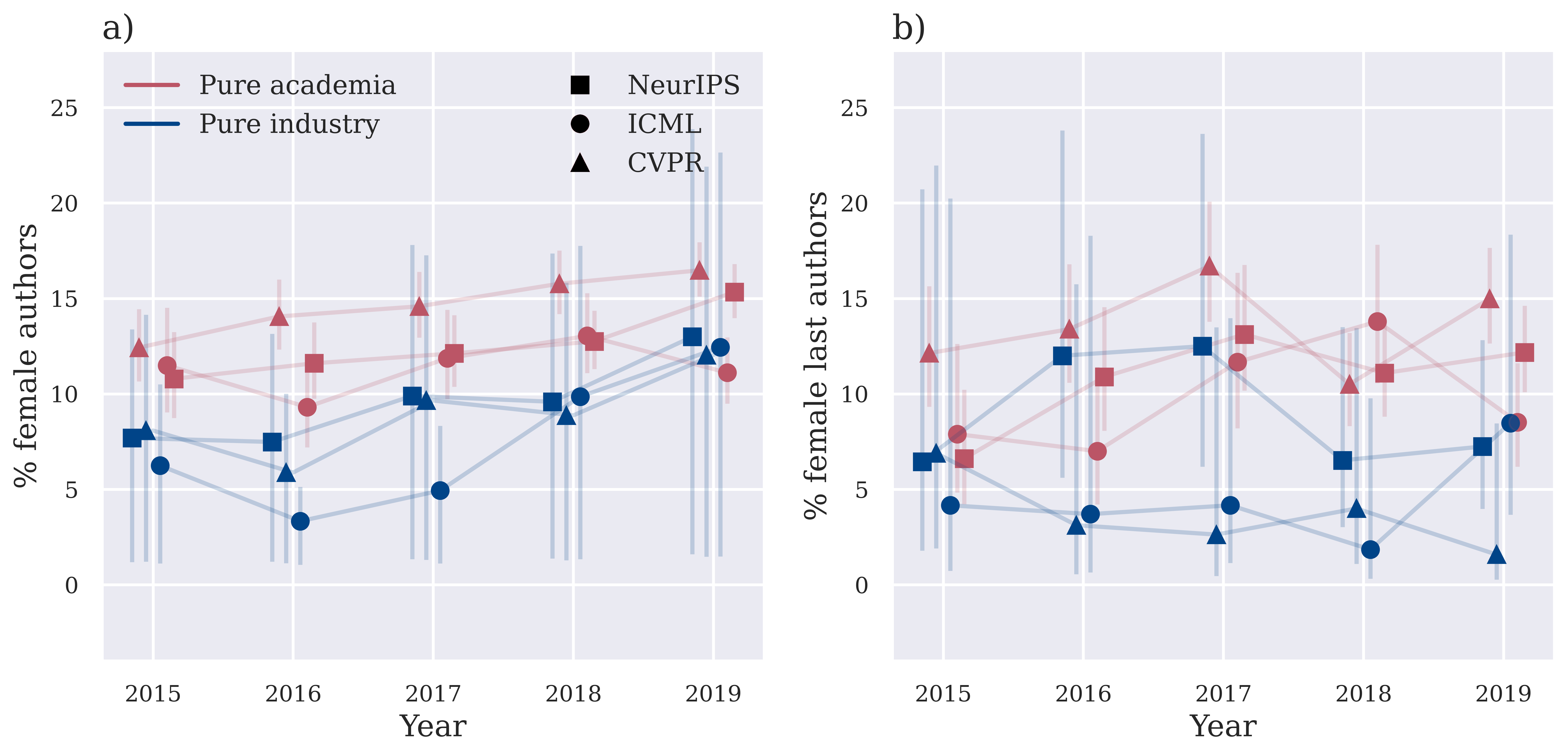}
		\caption{(a) Overall ratio of female authors in academia and industry and (b) ratio of last authors.}
		\label{fig:Fig4}
	\end{figure}

Finally, we analysed the contribution of male and female authors to ML conferences. We only focus on the difference between purely academic and purely industry papers since we are not able to assign the individual affiliations in mixed papers \label{sec::mixed}. Figure \ref{fig:Fig4}a shows the ratio of female to all authors across the conferences, indicating a slight increase in the ratio of female authors across the three major ML conferences.
\newpage
The ratio of female authors increases with 1.0\%/year (95\%-\(CI\): [0.7,1.3], \(p < 10^{-5}\)). It is somewhat noticeable, though, that female authors are represented less in industry papers, compared to academic papers. Our linear model analysis shows a difference of 4.1\% (95\%-\(CI\): \([3.0,5.3]\), \(p < 10^{-5}\)) between academia and industry. 

Albeit not a universal practice, being the last author typically corresponds to the principal investigator or the most senior author. Apart from that, papers may have multiple advising authors. Notwithstanding these exceptions, we assume that in most cases last authors are sole principal investigators. Taking up previous research \cite{Andersen.2019} and going into further detail, figure \ref{fig:Fig4}b shows the ratio of female last authors compared to all last authors. No significant trend in time is found here (0.4\%/year 95\%-\(CI\): \( [-0.1,0.9]\), \(p=0.10223\)), although our model analysis reveals that there is a difference of 7.3\% (95\%-\(CI\): \( [5.6,8.9]\), \(p = 10^{-5}\)) between industry and academic papers.\\
Taking up the results from our analysis, we see that female authors are less represented in industry papers than in academic papers. The results are in line with other studies, claiming that the proportion of women in ML research and in the number of workforces at major tech companies is typically hovering between 10 and 20 percent \cite{Yuan.2020}. Despite that, we are fully aware of the fact that 
gender equality is only one dimension in a broader spectrum of diversity \cite{Hopkins.1997}. It is 
obvious that other types of diversity like ethnicity (asian, black, hispanic, white, other), nationality, age, etc. could also be analysed with respect to their differences in industry and university contexts. But since it is not possible to reliably yield this information from our data set, we refrained from analysing other types of diversity besides gender (for further discussion on the reliability of gender extraction, see \ref{Authors and gender extraction}). 

\section{Conclusion}
The scientific success of ML research lured an increasing amount of industry partners to coalescence with academia. The growing number of papers stemming from academic-corporate collaboration is an indication of this. While medical journals require researchers to name conflicts of interest, the ML community slowly follows and obliges researchers to add transparency statements to their work. This seems reasonable, especially against the backdrop of an increasing number of academic-corporate collaborations and academic papers with industry acknowledgements. In this context, since 2020, NeurIPS, the largest ML conference in the world, requires researchers who make submissions to describe the potential broader impact their respective research has on society as well as to disclose financial conflicts of interest, that could directly or indirectly affect the research, in a funding transparency statement. Further efforts to introduce transparency declarations are to be welcomed, while at the same time a responsible interpretation of these declarations is required in order to ensure that disclosure brings about the intended effects \cite{Loewenstein.2012}.\\
Up to now, though, only a handful of papers voluntarily add conflicts of interest sections. On a related note, it is difficult to describe concrete ramifications on lines of action, opinions, or advice. In medical research, tangible and relatively direct influences from the pharmaceutical industry can be picked up. In ML research, industry influences are more fuzzier and hard to monitor. Hence, the concrete consequences of existing conflicts of interest can only be discovered by more in-depth, qualitative empirical social research. One can assume that in ML research ramifications mostly affect research agendas, so that scientists consciously or unconsciously steer their research in a direction that is most valuable for corporate interests or commercialization processes of all kinds. This bias can also suppress certain research results in order to avoid unfavorable outcomes that are nonpractical to those interests or processes. After all, universities and companies follow different ``symbolically generalized communication media'' (i.e. money or truth, see \cite{Luhmann.1995b}), which can make it difficult for researchers with corporate cooperation to act in accordance with only one of those goals. In this context, it is important to keep in mind that even small gifts or favors elicit the reciprocity principle, meaning that individual behavior is under the influence of an industry bias.\\
Despite the issue of conflicting interest, our data analysis provides evidence for the fact that industry-driven research has a measurable impact and is setting research trends ahead of academia. This insight stands in contrast to the rather industry-critical discourse on conflicts of interest and proves the irrefutable positive impact industry-driven research has on scientific progress in ML. In line with this insight, we show that industry papers receive significantly more citations than research from academia, which is a clear sign that corporate ML research is of high importance for the scientific community. Besides the great attention that is directed towards industry papers, we demonstrate that these papers are not just orientated towards technical issues and do not omit to discuss social aspects of technology, as one might be tempted to impute. Actually, the amount of social impact terms that we used to measure the significance of social aspects is more or less equally distributed between academic and industry papers. This finding hints to the fact that something like an ethical ``superiority'' of academia against industry does not necessarily exist.\\
Tangible problems, however, occur in view of diversity shortcomings. We show that the ratio of female authors compared to male authors of conference papers indicates a slight improvement of gender equality over time. But overall, the proportion of women in ML research is quite small. This holds especially true with respect to industry research. Here, amendments are necessary, mainly comprising the creation of more inclusive workplaces, changes in hiring practices, but also an end of pay and opportunity inequalities \cite{Crawford.2019}. In contrast to issues like innovative strength or citations, industry has a lot of catching up to do here.\\
In summary, we provide quantitative evidence for the increasing influence tech companies have on ML research. Our analysis reveals three main insights that can inform and differentiate future policies and principles of research ethics. Firstly, the analysis shows that besides the growing number of academic-corporate collaborations, conflicts of interest are not disclosed sufficiently. Secondly, it proves that industry-led papers are not only a strong driving force for promising scientific methods, but possess significantly more citations than academic papers, while being in no way inferior with regard to considerations on social impacts. Thirdly, we provide further evidence for the need to improve gender balance in ML research, especially in industry contexts.

\section*{Acknowledgements}
This research was supported by the Cluster of Excellence “Machine Learning – New Perspectives for Science” funded by the Deutsche Forschungsgemeinschaft (DFG, German Research Foundation) under Germany’s Excellence Strategy – Reference Number EXC 2064/1 – Project ID 390727645. 
Additional funding was provided for K.M, in part, by the Deutsche Forschungsgemeinschaft (DFG, German Research Foundation) -- projectnumber 276693517 -- SFB 1233, TP 4 Causal inference strategies in human vision. \\
We would like to thank Felix Wichmann, Ulrike von Luxburg, Sarah Fabi, and Cornelius Schröder for very helpful comments on the manuscript, and Ella Gierss as well as Daniela Gjuraj for helping us with the preparation of the manuscript. 
	
\section*{Author Contributions}
Both Kristof Meding and Thilo Hagendorff developed the ideas for this paper. Kristof Meding performed the statistical analysis of the papers and did the computations, wrote the Python script, and composed the figures with input from Thilo Hagendorff. Thilo Hagendorff wrote large parts of the manuscript, developing the theory as well as the ethical analysis. All authors discussed the results and contributed equally to the final manuscript.

\section*{Competing Interests statement}
The authors declare that they have no conflict of interest.

	\bibliographystyle{naturemag}
	\bibliography{Literature.bib}
	\newpage
	\appendix
	\counterwithin{figure}{section}
	\renewcommand{\thesection}{} 
	\renewcommand{\thesubsection}{S.\arabic{subsection}} 
	\counterwithin{figure}{section}
	\section{Supplementary Material} 
	Section \ref{DataAnalysis} provides in-depth explanation and all information necessary for reproducing our results. Section \ref{sec:FiguresAppendix} shows additional figures.
	\renewcommand{\thesection}{SF} 

	\subsection{Data Analysis}
	\label{DataAnalysis}
	\subsubsection{Code Availability}
	Unfortunately, due to legal restrictions, we can neither offer our database of papers nor our script for downloading the papers to the public. Our analysis is not usable without this database.
	However, we are looking forward to receive any questions regarding our approach and we are happy to answer all inquiries to ensure reproducibility.
    
	\subsubsection{Paper download}
	In total, we have downloaded 10,807 papers.
	The NeurIPS papers were downloaded from \url{https://papers.nips.cc/} and ICML/CVPR from \url{http://proceedings.mlr.press/}. To avoid traffic overload on servers, we recommend a minimum delay between each individual paper download.  
	Especially the automatic download of NeurIPS papers sometimes fails. This problem can be caused by the download link having a different name than the paper or by the download link being too long for our Linux file system (Ubuntu 18.04 LTS 64 bit). Therefore, every time the download failed, we manually added the corresponding paper to our collection. This is also necessary for reproducing our results.
	
	\subsubsection{Text conversion}
	It is a challenge to convert pdfs to text, since the PDF format is not suitable or meant for this task. We have decided to use the command line tool pdf2text.The tool pdf2text is able to convert pdf-documents into simpler txt-documents. We made sure that 2-column design is converted correctly. As a further pre-processing step we removed the watermark and headers of the ICML papers. We preserved the bibliography in each case, as this also gives an insight into the content. The supplementary materials, on the other hand, remained neglected. \\
	With this procedure, we were able to obtain 10,772 papers which contain at least the word ``the''. The outlying papers are e.g. those in which the entire text was present as an image.
	
	\subsubsection{Text search}
	We came up with a simple function that searches the text files. This function was not case-sensitive and finds arbitrary subwords, it can for example find the word ``anonym'' within the text segment ``anonymous reviewers''. With blanks, the search can be limited to certain words. To avoid unintentional results, we have compared all terms we have searched for in the data set with the following list of English words: \url{https://raw.githubusercontent.com/dwyl/english-words/master/words.txt}
	For the search for social impact words we used the list on the corresponding NeurIPS 2020 subject area. However, we removed some words to improve the results. Caution is necessary here, as one might receive false positive results when, for example, ``anonymity'' is changed to ``anonym'' as many authors thank their anonymous reviewers. Finally, we used the terms ``AI safety'', ``fairn'', ``accountab'', ``transpar'', ``privacy'', ``anonymity'', ``security'' and the terms ``ethic'' and ``explainab''.
	
	\subsubsection{Affiliation extraction}
	We are not only interested in analyzing the content of the papers, but also their origin. Therefore, we have tried to extract the headers of the papers. This was no problem for NeurIPS or CVPR papers. For these papers, we simply extracted all content before the word ``abstract''. In most cases, there were no issues. Very rarely, a figure appeared before the abstract or authors changed the standard template. The same procedure worked for ICML 2015 and 2016. However, from 2017 onwards, the affiliations were shown in the lower left corner. No keywords were placed before, only a blank line. This was difficult to parse with our script. We thus decided to keep the first 5000 characters as header for these papers, but split it before the terms ``international conference of machine learning'' which always ended the listing of authors. We think that this yields only a small amount of false positive if we search for affiliations, since it is most likely that the academic and industry institutional terms will appear in the affiliations only. \\
	To get an impression of which institutions publish on NeurIPS, CVPR, and ICML, we followed preexisting analyses:
	\begin{itemize}
		\item \url{https://www.microsoft.com/en-us/research/project/academic/articles/neurips-conference-analytics/},
		\item \url{https://www.reddit.com/r/MachineLearning/comments/bn82ze/n_icml_2019_accepted_paper_stats/}
		\item \url{https://medium.com/@dcharrezt/neurips-2019-stats-c91346d31c8f}
		\item \url{https://www.microsoft.com/en-us/research/project/academic/articles/eccv-conference-analytics/}
	\end{itemize}
	To prevent us from cherry-picking, we only used terms appearing in the analyses above.
	We define a paper as academic if the affiliation section contains one of the following terms:\\ \label{sec::affTerms}
	\\
	California Institute of Technology
	/ Ecole
	/ EPFL
	/ ETH Z\"urich
	/ INRIA
	/ Kaist
	/ Massachusetts Institute of Technology
	/ MILA
	/ MIT
	/ ParisTech
	/ Planck
	/ RIKEN
	/ TU Darmstadt
	/ Université
	/ Universiteit
	/ University. \\
	\\
	For the definition of a paper as industry we use the following terms: \\
	\\
	Adobe
	/ AITRICS
	/ Alibaba
	/ Amazon
	/ Ant Financial
	/ Apple
	/ Bell Labs
	/ Bosch
	/ Criteo
	/\linebreak Data61
	/ DeepMind
	/ Expedia
	/ Facebook
	/ Google
	/ Huawei
	/ IBM
	/ Intel
	/ Kwai
	/ Microsoft
	/ NEC
	/ Netflix
	/ NTT
	/ Nvidia.
	/ Petuum
	/ Qualcomm
	/ Salesforce
	/ Siemens
	/ Tencent
	/ Toyota
	/ Uber
	/ Vector Institute
	/ Xerox
	/ Yahoo
	/ Yandex.\\
	\\
	We perform a non-exclusive classification. Papers may have academic and industry affiliations. It is important to note that we included blanks before and after the text for the MIT, NEC and Intel terms to avoid contaminations with other words like ``admit''. 
	\subsubsection{Extract acknowledgements}
	We extracted the acknowledgements for our conflict of interest analysis. In this particular analysis, we focused on academic papers. In our data sets, we have 6632 papers from academia. Out of these papers, 5221 papers (78.7\%) contain an acknowledgement section which we were able to parse. We also included both spellings of acknowledgement: ``acknowledgement'' and ``acknowledgment''.
	
	\subsubsection{Authors and gender extraction} \label{Authors and gender extraction}
	The authors were not imported from the PDFs, but from the websites. We found a total of 41,939 authors. However, it is clear that some papers were written by the same author. Therefore, we decided to pool all authors with the same name. Of course, this leads to the effect that different authors with the same name are pooled. We believe that this effect is negligible. 
	For authors with middle names, we kept only the first letter. People vary the ways in which they indicate their middle name, e.g. T., T, or Tom. This procedure gives us 18,060 unique authors.
	From these unique authors, we extracted the gender using the commercial tool GenderAPI. GenderAPI also provides an estimate of the accuracy. The mean accuracy in our case was 87.1\%. Unfortunately, we noticed that most times, GenderAPI fails in the recognition of names from Asian language families. This is a clear bias in the underlying dataset of GenderAPI. Furthermore, we want to acknowledge that some people reject the idea that a name corresponds to a gender. However, we applied the analysis of genders here to gain insight into the inequality of author's genders on average, not only in single cases. 
	
	\subsubsection{Downloading citations}
	Finally, we conducted a citation analysis. To do this, we first downloaded the titles of the papers from the websites with Beautiful Soap. We then wrote an automated script to access the Microsoft Academic Knowledge API \cite{Sinha2015}. This was successful for 10,616 papers (98.2\%, date of citation download: 03.29.2020). The most common reason for a paper not being found in the database is the use of special characters like \(\ell,\lambda,\) etc. in the title.
		
	\newpage
	\subsection{Additional Figures}
	\label{sec:FiguresAppendix}
	
	\subsubsection{Figure A1}
	\label{sec::FiguresStart}
	\begin{figure}[!h]
		\centering
		\includegraphics[width=1\textwidth]{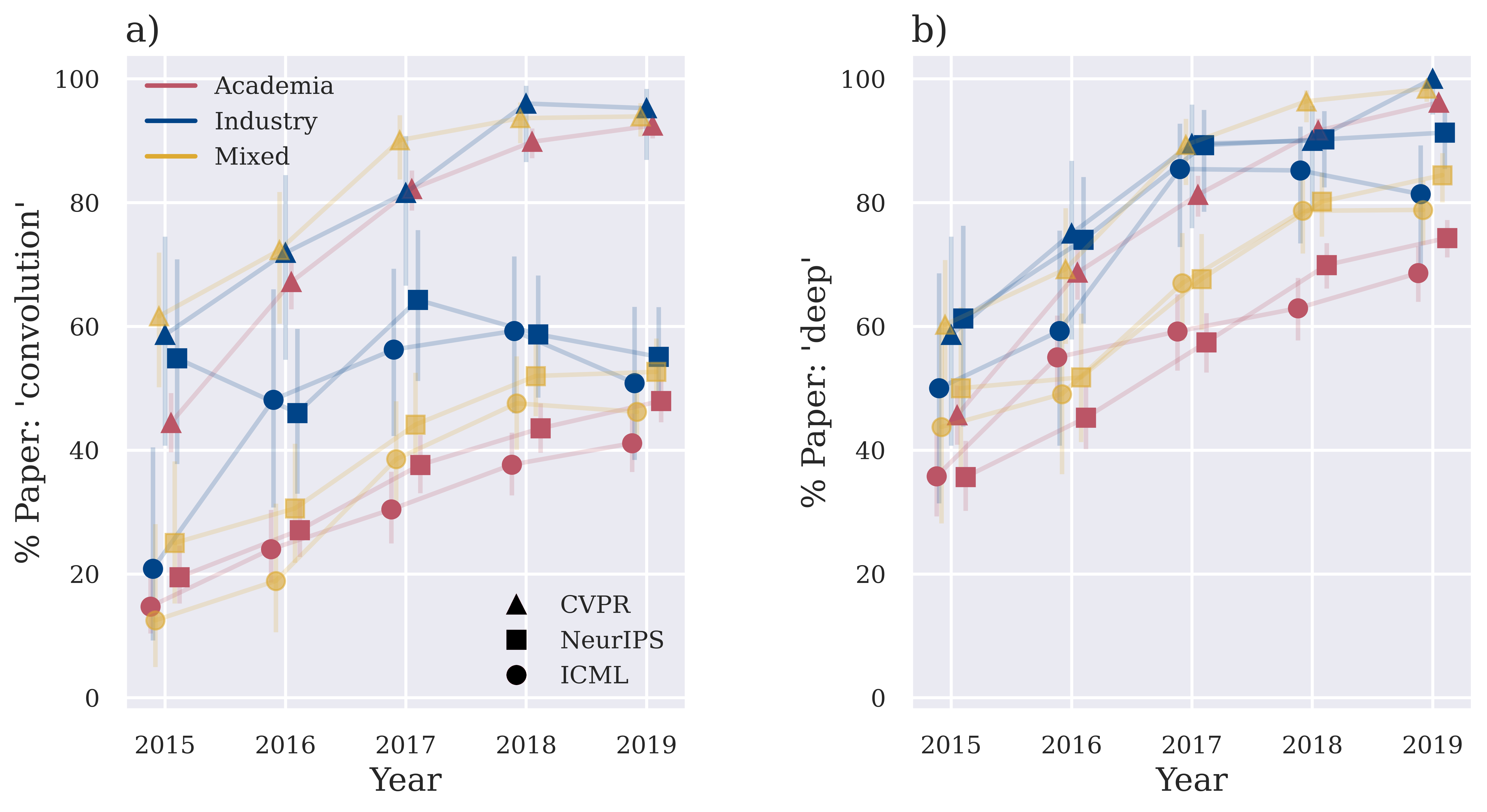}
		\caption{Ratio of academic (red), mixed (orange) and industry (blue) paper which include the term ``convolution'' (a) or ``deep'' (b).}
		\label{fig:FigA1}
	\end{figure}
	
	\newpage

	\newpage
	\subsection{Model details}
	\label{sec:ModelDetails}
	\subsubsection{A note on statistical modelling}
    We fitted the data of the figures directly for almost all figures. We included the standard error of the individual datapoints from the figures with the \textit{metafor} package in R for our analysis. 
	Only in case of the citation counts we model the citation count of the individual papers as the dependent variable, since every single paper contains a citation count. This also corresponds to figure \ref{fig:Fig3}a.
	 In all models the variables \textit{conference} and \text{affiliation} were used as a factorial variable. The reference level of the conference variable was \textit{CVPR}. The reference level for the affiliations variable was \textit{Academia}.
		
	\subsubsection{Figure 1b}
	We modelled the data of figure 1b as linear combination of the conference and year. Separate models are fitted for academic, industry, and mixed papers since the effect of year is increasing for mixed papers and decreasing for academia papers. Please note that the large negative intercept comes from the fact that the year starts with 2015. \\
\\	Pure academia:
\begin{lstlisting}[language=R, firstline=21, lastline=28]
                   estimate       se      zval     pval     ci.lb     ci.ub 
intrcpt            59.36487  8.39702   7.06975  <.00001  42.90700  75.82273 
Year               -0.02910  0.00416  -6.99134  <.00001  -0.03726  -0.02094 
ConferenceICML     -0.07278  0.01493  -4.87477  <.00001  -0.10204  -0.04352 
ConferenceNeurIPS  -0.05190  0.01351  -3.84104  0.00012  -0.07838  -0.02542 

\end{lstlisting}
	Pure industry:
\begin{lstlisting}[language=R, firstline=21, lastline=28]
                   estimate       se      zval     pval     ci.lb    ci.ub 
intrcpt            -1.26103  3.48674  -0.36166  0.71760  -8.09492  5.57285 
Year                0.00065  0.00173   0.37580  0.70707  -0.00274  0.00404 
ConferenceICML      0.03693  0.00660   5.59441  <.00001   0.02399  0.04987 
ConferenceNeurIPS   0.04001  0.00555   7.20410  <.00001   0.02913  0.05090 

\end{lstlisting}
	\newpage
\noindent	Mixed:
\begin{lstlisting}[language=R, firstline=21, lastline=28]
                    estimate       se      zval     pval      ci.lb      ci.ub 
intrcpt            -71.98876  8.60753  -8.36346  <.00001  -88.85922  -55.11831 
Year                 0.03578  0.00427   8.38334  <.00001    0.02741    0.04414 
ConferenceICML       0.05066  0.01533   3.30363  0.00095    0.02060    0.08072 
ConferenceNeurIPS    0.01356  0.01426   0.95089  0.34166   -0.01439    0.04150 

\end{lstlisting}
	
	\subsubsection{Figure 2}
	The model for figure 2 is similar to the model of figure 1b.
\begin{lstlisting}[language=R, firstline=21, lastline=28]
                   estimate       se      zval     pval      ci.lb    ci.ub 
intrcpt            -9.60675  9.91590  -0.96882  0.33263  -29.04156  9.82805 
Year                0.00485  0.00492   0.98664  0.32382   -0.00479  0.01449 
ConferenceICML     -0.01597  0.01714  -0.93214  0.35126   -0.04956  0.01761 
ConferenceNeurIPS  -0.01292  0.01718  -0.75217  0.45195   -0.04658  0.02074 

\end{lstlisting}
	
	\subsubsection{Figure 3a}
	Here, we directly modelled the effect of citation count of every single paper. We tested a Poisson and a negative binomial distribution. Finally, we used a negative binomial distribution since this distribution leads to a lower Bayesian Information Criterion and more detailed deviance compared to the Poisson distribution. 
\begin{lstlisting}[language=R, firstline=1, lastline=17]

Call:
glm.nb(formula = CitationCount ~ Year + Conference + Type, data = DataFigure3a, 
    init.theta = 0.4963353512, link = log)

Deviance Residuals: 
    Min       1Q   Median       3Q      Max  
-2.4750  -1.1320  -0.6233  -0.0250  13.4232  

Coefficients:
                    Estimate Std. Error z value Pr(>|z|)
(Intercept)       1494.40994   22.12978   67.53   <2e-16
Year                -0.73907    0.01097  -67.37   <2e-16
ConferenceICML      -0.70696    0.03914  -18.06   <2e-16
ConferenceNeurIPS   -0.98047    0.03406  -28.79   <2e-16
TypeIndustry         0.98497    0.05457   18.05   <2e-16
TypeMixed            0.54534    0.03612   15.10   <2e-16
\end{lstlisting}
	
	\newpage
	\subsubsection{Figure 3b and figure 3c}
	For figure 3b and figure 3c, we again modelled the data of the figure as in figure 1b and figure 2. This time we can include the authorship affiliation since we observe the same trend for academic, industry, and mixed papers.
\begin{lstlisting}[language=R,basicstyle=\tiny, firstline=16, lastline=29]
                     estimate        se       zval     pval       ci.lb       ci.ub 
intrcpt            -230.12876  17.65614  -13.03392  <.00001  -264.73417  -195.52336 
Year                  0.11419   0.00875   13.04385  <.00001     0.09703     0.13135 
ConferenceICML        0.15158   0.03068    4.94016  <.00001     0.09144     0.21171 
ConferenceNeurIPS     0.16907   0.02978    5.67766  <.00001     0.11071     0.22744 
TypeIndustry          0.11515   0.03133    3.67575  0.00024     0.05375     0.17655 
TypeMixed             0.07043   0.02897    2.43097  0.01506     0.01365     0.12721 

\end{lstlisting}
\begin{lstlisting}[language=R,basicstyle=\tiny, firstline=21, lastline=30]
                    estimate       se      zval     pval      ci.lb      ci.ub 
intrcpt            -65.23549  7.58347  -8.60232  <.00001  -80.09882  -50.37215 
Year                 0.03241  0.00376   8.62113  <.00001    0.02504    0.03978 
ConferenceICML       0.01735  0.01324   1.31032  0.19009   -0.00860    0.04330 
ConferenceNeurIPS    0.01395  0.01227   1.13695  0.25556   -0.01010    0.03801 
TypeIndustry        -0.03110  0.01467  -2.12054  0.03396   -0.05985   -0.00236 
TypeMixed            0.00036  0.01213   0.02926  0.97666   -0.02343    0.02414 

\end{lstlisting}
	
	\subsubsection{Figure 4a and figure 4b}
	The reasoning of figure 4a and figure 4b is the same as in figure 3b and figure 3c. We only model the difference between industry and academia and exclude mixed papers, see the explanation in section \ref{sec::mixed}.\\
	Figure 4a:\\
\begin{lstlisting}[language=R,basicstyle=\tiny, firstline=18, lastline=27]
                      estimate       se      zval     pval      ci.lb      ci.ub 
intrcpt              -19.36623  3.19982  -6.05229  <.00001  -25.63776  -13.09470 
Year                   0.00967  0.00159   6.09832  <.00001    0.00656    0.01278 
ConferenceICML        -0.03468  0.00570  -6.08202  <.00001   -0.04585   -0.02350 
ConferenceNeurIPS     -0.01684  0.00511  -3.29610  0.00098   -0.02685   -0.00683 
AffiliationIndustry   -0.04131  0.00591  -6.99407  <.00001   -0.05289   -0.02974 

\end{lstlisting}
	 Figure 4b \\
\begin{lstlisting}[language=R, basicstyle=\tiny,firstline=18, lastline=32]
                     estimate       se      zval     pval      ci.lb     ci.ub 
intrcpt              -7.86349  4.88807  -1.60871  0.10768  -17.44393   1.71695 
Year                  0.00396  0.00242   1.63414  0.10223   -0.00079   0.00871 
ConferenceICML       -0.02544  0.00867  -2.93457  0.00334   -0.04243  -0.00845 
ConferenceNeurIPS    -0.00854  0.00801  -1.06549  0.28665   -0.02424   0.00717 
AffiliationIndustry  -0.07262  0.00824  -8.81558  <.00001   -0.08876  -0.05647 

\end{lstlisting}
\end{document}